# An Efficient Reconfigurable FIR Digital Filter Using Modified Distribute Arithmetic Technique

Naveen Shankar Naik[1], Dr. Kiran Gupta, Senior Member IEEE[2]

[1,2]*Dept. of Electronics and Communication, DSCE Bangalore, India*

*Abstract--* **In various telecommunication applications Digital Signal Processors are the key components in transferring the data between devices. The implementation of FIR filter on FPGA is based conventional methods increasing the need for considerable hardware resources, which in turn raises the circuit size and lowers the system speed. Most important operation performed in digital signal processing is Multiply and Accumulation (MAC). Usually this operation is realized using novel hardware multipliers. The computations for sum of products can be performed more effectively using Distributed Arithmetic. This paper provides modified Distributed Arithmetic based technique to compute sum of products saving appreciable number of Multiply And accumulation blocks and this consecutively reduces circuit size. In this technique multiplexer based structure is used to reuse the blocks so as to reduce the required memory locations. In this technique a Carry Look Ahead based adder tree is used to have better area-delay product. Designing of FIR filter is done using VHDL and synthesized using Xilinx 12.2 synthesis tool and ISIM simulator. The power analysis is done using Xilinx Xpower analyzer. The proposed structure requires nearly 42% less cells, 40% less LUT flip-flop pairs used, and also 2% less power compared with existing structure.**

*Keywords*—**Distributed arithmetic, Look up Table, Finite Impulse Response Filter**

## I. INTRODUCTION

Signal processing is one of the most demanding applications in both wired and wireless networks with enhanced performance. In recent years many telephony and data communications are moving to digital domain because they attain much better signal to noise ratios than analog processing domain. Therefore the need for efficient digital filtering methods continues to grow[4].

There is raising trend to implement Digital Signal Processing in Field Programmable Gate Arrays (FPGA's). Along with advancement of Very Large Scale Integrated Circuits (VLSI), Digital signal processing has become increasingly popular in recent years. In the sense, we need to put more effort in designing efficient architecture for DSP functions such as FIR filters, which are broadly used in video and audio signal processing of telecommunication [5].

Since the complexity and circuit size of higher order filters are too large, the real time processing of these filters with desired area and power metrics is challenging task [3].

Traditional FIR filters require Multiply and Accumulation (MAC) blocks which are expensive to implement. To avoid this problem, we present Distributed Arithmetic technique serving as multiplier-less architecture for digital signal processing applications. In this paper we are optimizing this technique in terms of area-delay product [3]. To get overall performance and to minimize the access delay and power dissipation either the processor has been move to memory or memory has been moved to processor. since semiconductors are scaled down to tremendous value, memory based architectures are well suited for many DSP applications [3]. Memory components like ROM and RAM are utilized as part of or whole arithmetic structures. Memory based architectures are regular to the greater extent, has very high potential, reduced circuit size and less dynamic power consumption compared to conventional multipliers. One of the memory based technique is Distributed arithmetic whose co-efficients are transformed to another numeric representation where the arithmetic manipulation is more efficient than traditional implementation [5].

## II. DISTRIBUTED ARITHMETIC (DA)

Filters are usually frequency selective networks, capable of modifying an input signal in order to facilitate further processing. Usually digital filters are more preferred than analog due to its high signal integrity [2]. FIR filters are widely applied for variety of DSP areas because providing virtues of linear phase and system stability. The basic convolution equations of filter representation are shown as follows

$$y(n) = b_0 x(n) + b_1 x(n-1) + \cdots + b_N x(n-N) \quad (1)$$

$$= \sum_{i=0}^{N} b_i x(n-i) \quad (2)$$

Distributed Arithmetic is termed so because the multiplications that appear in signal processing are reordered and combined such that the arithmetic becomes distributed completely through the structure rather than being lumped. Multipliers are replaced by combinational Look Up Tables (LUT) [3]. Since LUTs are considerably larger in size, the quality of implementing FIR filter mainly relies upon the efficiency of logic synthesis algorithm mapping to FPGA.

DA provides bit serial operations that implement a series of fixed point MAC operations in a known number of steps, regardless of number of terms to be calculated. The main operations required for DA based computation of inner product are sequence of lookup table access followed by shift accumulation operations of LUT output [9].

According to DA, we can make look up tables (LUT) to store MAC values and call out values accordingly to the input data if necessary. Therefore, LUT's are utilized to facilitate the operations of MAC units so as to save hardware resources. This technique also facilitates DA computation suitable for FPGA realization, because the LUT along with shift and add operations can be directly mapped to LUT base FPGA logic structure.

The following expressions represent multiply and accumulate operation [8]

$$i.e \quad Y = \sum_{k=1}^{K} A_k x_k \quad (3)$$

$$x_k = -b_{k0} + \sum_{k=1}^{K} b_{kn} 2^{-n} \quad (4)$$

$$Y = \sum_{k=1}^{K} A_k [-b_{k0} + \sum_{k=1}^{K} b_{kn} 2^{-n}] \quad (5)$$

$$Y = -\sum_{k=1}^{K} A_k b_{k0} + \sum_{k=1}^{K} [\sum_{n=1}^{N-1}(A_k b_{kn}) 2^{-n}] \quad (6)$$

The final formulation is

$$Y = -\sum_{k=1}^{K} A_k b_{k0} + \sum_{n=1}^{N-1} [\sum_{k=1}^{K}(A_k b_{kn}) 2^{-n}] \quad (7)$$

The advantage of DA is its efficiency of mechanization. It turns out well when the number of elements in a vector is nearly same as word size then DA is quite fast.

### III. FIR FILTER WITH DA

Distributed arithmetic is well known method of implementing FIR filters without the use of multipliers. In DA the task of summing product terms is replaced by table lookup procedures that are easily implemented on FPGA.

In FIR filtering, one to one mapping of the convolving sequences are from two sources, one obtained from input samples while the other sequences are obtained from fixed impulse response coefficients of filter [6]. This behavior of FIR filter makes it suitable for DA based technique for memory realization. It yields faster output, it stores pre-computed partial results in memory components that can be read out and accumulated to obtain desired result [8].

The Distributed Arithmetic technique of FIR filter consists of Look-Up Table (LUT), Shifters and accumulator with adder tree. In this technique all the partial product outputs are pre-computed and placed in a Look Up Table (LUT). These table entries are addressed by the addresses generated by input multiplier bit data. All inputs are fed simultaneously. From the input data, address are generate and allowed to access the LUT; its outcome is added to the accumulated partial products.

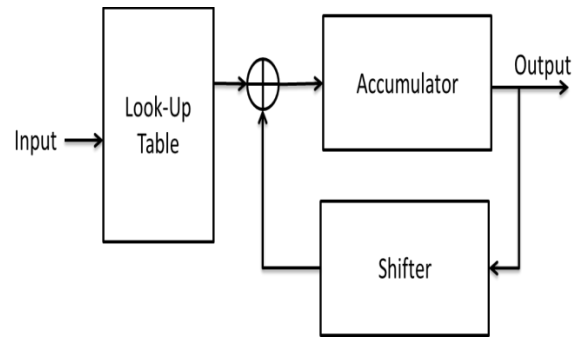

**Figure1: Block diagram for FIR filter using DA**

The Basic block diagram for FIR filter implementation using DA [8] is shown in Figure 1. The complete dot product calculations takes L clocks where L is the size of input data, and it is not depended on input data size. During the initial cycle of operation, the Least-Significant Bits of input i.e., $X_0(n)$, $X_0(n-1)…$, of the K input samples are arranged to form K-bit addresses. These addresses are allowed to access Look-Up Table and that table outcome becomes the initial value of the accumulator. During the very next cycle of operation, the next-to-least significant bits $X_1(n)$, $X_1(n-1)$, ..., $X_1(n-K+1)$ of the K input samples are arranged to form K-bit address which will be allowed access lookup, and the adder sums the Look up Table output is shifted by one bit and summed to the contents of the accumulator.

This process continues until the last addresses are allowed to access Look up table, that is, the most-significant bits $X_{N-1}(n)$, $X_{N-1}(n-1)$,..., $X_{N-1}(n-K+1)$ of the K input samples form an K-bit address are arranged to form K-bit address which will be allowed access lookup and the Look up Table outcomes are added to the contents of the accumulator after shifting it to the corresponding position and this will be stored it in accumulator. A filter with N coefficients Look-Up Table requires $2^N$ values [4]. Larger the value of N, the LUT size will grow very high and it acquires more storage space. This consecutively lowers the performance and turns as bottleneck of the design.

### IV. LUT PARTITIONING

The size of LUT increments exponentially with the input data bits of filter. In order to have greater performance for higher order filters, LUT size should be minimized to sensible levels. To lower the total size, the LUT can be divided into many LUTs, called LUT partitioning [8]. The Figure 2 gives the idea of FIR filter realization with 4 input channels each are 4 bits wide. In this technique the LUTs are partitioned into two separate LUTs. Each LUT constitutes of 4 memory locations. Each LUT partition operation is performed on a set of different filter taps. The outputs obtained from the partitions are summed to obtain final output.

For 3rd order filter
Number of partitions = 2
Number of Look-up tables used=2 and Each look-up table has 2 inputs as address.
So total Memory locations = Number of partitions * $2^N$ = $2*2^2$ = 8 locations.
N = size of inputs of LUT

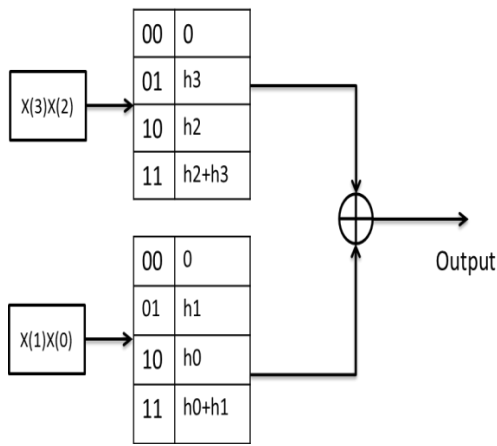

**Figure 2: LUT table for 3 tap FIR filter with LUT partitioning**

The LUT partitioned method will require 8 memory locations, whereas the conventional method requires 16 memory locations for 3 tap filters [8].

### V. PROPOSED ARCHITECTURE

FPGA technology has grown rapidly from dedicated hardware to a dissimilar system which is considered as a famous choice in communication. The proposed architecture of FIR filter suggested in this section gives an area efficient implementation FIR filter where the area-delay product can also be reduced.

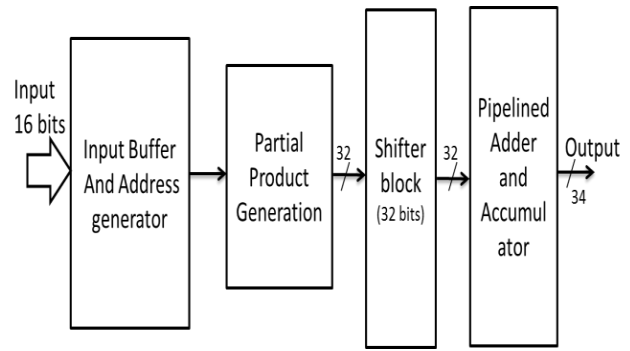

**Figure3: Block Diagram of proposed FIR filter implementation**

Figure 3 shows the complete overview of the implementation of FIR filter using the proposed share memory based structure. Each block is synchronized using a common clock. Every two bits address are allowed to generate the partial products at a time and the results shifted using shifter block based on the address bit precision and accumulated using the accumulator. The output will be 34 bits wide this is because, 8 parallel inputs data is fed to the filter and the addition of 8 parallel filters needs one bit extra. MSB represents sign bit. Thus, total 34 bits are required at the accumulator.

The basic block diagram consists of sub modules such as: Address Generator, Partial product Generator, Pipelined adder tree.

*A. Input buffer:*

The inputs are fed to input buffer in parallel manner in the form of bus. The addresses are generated from the input multiplier bits on the arrival of each clock edge. The generated addresses are stored in the input buffer. In a pipelined way the addresses are allowed to generate the partial products.

*B. Partial product generation*

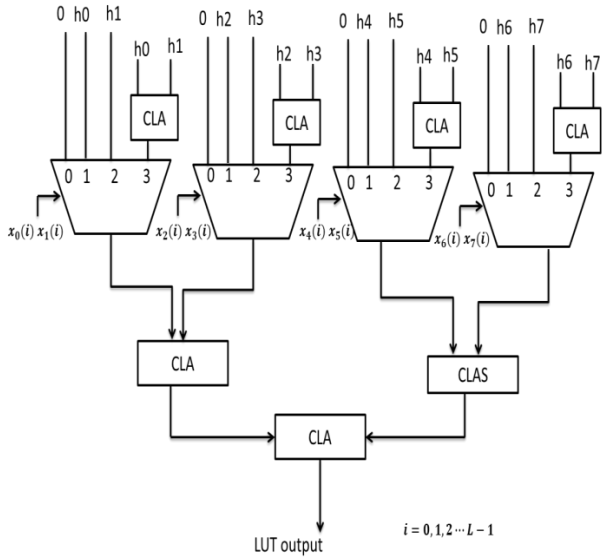

**Figure 4: Detailed Block Diagram for LUT access for the proposed FIR filter**

Figure 4 shows the implementation of partial product generation for 8 inputs. The partial product generation block is a share memory based structure [1]. In this architecture partial product generation is done using multiplexer structures and modified carry look-ahead adder. In the following implementation, the filter coefficients and the addresses generated by the multiplier bits are fed to the Partial product generation (PPG) block. The partial product generation block takes 8 addresses each of which 2 bits wide. The in-place computation of partial product is done using shared memory based method. Each filter coefficients are 16 bits wide. So the output of each multiplexer are 17 bits wide. When all the multiplexer outputs are added together the size will be 18 bits wide. This partial product generation block is repeatedly used for all the other addresses. The outcome of PPG unit will then be fed to accumulation and shifting unit.

*C. Accumulator and shifter unit*

This stage includes an accumulator, adder and a shifter. On arrival of each clock cycle, the partial products generated are shifted which serves as multiplier and added using modified Carry Look Ahead adder based adder tree. Number of bit shifting will be decided by the input address bit precision.

## VI. RESULTS AND DISCUSSION

The implementation results of 34-tap FIR filter are obtained by applying distributed arithmetic algorithm as shown in Table 1 and synthesized using Xilinx ISE 12.2, Simulated using ISIM simulator and results are compared with the existing CSA based structure. The designed FIR filter is programmed in VHDL language. The FIR filter is programmed for CSA based structure as well as CLA based structure for performance analysis. Its obtained as CLA based structure is superior to CSA based structure in many of the performance metrics as shown in Table 1. The Table shows the comparison of parameters of CLA based structure for FIR implementation over the existing CSA based structure for K =8 and L= 16. The area delay product can be obtained as follows

Area-delay product (ADP) = Cells Used* Minimum Time

ADP for CSA structure = 606*2.375ns

= 1439.25 cell-ns

ADP for CLA structure = 357*2.523ns

=900.71 cell-ns

The number of cells used, number of slice LUTs (SLUT), LUT flip-flop pairs used are listed in Table 1. The proposed structure requires nearly 42% less cells, 40% less LUT flip-flop pairs used, and also 2% less power compared with CSA based structure. Using the modified CLA based FIR filter, area-delay product have been reduced to 37% when it's compared with CSA based structure.

Even though the area is reduced to 40% there is not enough reduction in power this is because CLA structure consumes more power compared to that of existing CSA structure and the parallel processing increases the power due to more number of computation. But reusing Partial Product Generation block to L number of times, there is significant reduction in area.

**Table1:**
**Parameter comparison table for CSA based existing architecture and CLA based enhanced architecture**

| Parameters | Existing (CSA based structure) | Enhanced (CLA based structure) | Improvement |
|---|---|---|---|
| Cells Used | 606 | 357 | 42% less cells |
| Slice LUTs | 567 (2%) | 356 (1%) | 1% of total LUTs |
| LUT- Flip-Flops pairs | 656 | 401 | 40% LESS flip-flops |
| Power | 387mW | 379mW | 2% less power |
| Minimum time | 2.375ns | 2.523ns | 7 % more |

## VII. CONCLUSION

The proposed architecture shows the area efficient implementation of FIR filter. This architecture replaces the complicated multiplication-accumulation operation with simple shifting and adding operations with the shared memory based DA algorithm which is directly applied to realize FIR filter. This project reports the improved shared DA architectures for high-order FIR filter. This architecture reduces the memory usage by 40% less LUT-flip-flop pairs at the cost of decrease in maximum frequency. And we used CLA based structure for addition purpose which serves as the pipelined adder tree. This technique reduces power, area-delay product and also the performance is improved by pipelining all the partial tables. This architecture requires nearly 42% less cells, 40% less LUT flip-flop pairs, 37% and also 2% less power compared with CSA based existing structure which can be seen is Table1.

The area-delay product has also been reduced to 37%. The performance of LUT based FIR structures with different adders can be studied in future. The performance can be further improved by pipelining all the partial tables.